\newcommand{\ttt}{t\bar{t}}
\newcommand{\alphas}{\alpha_{s}}
\newcommand{\afb}{A_{FB}}
\newcommand{\alab}{A_{FB}^{p\bar{p}}}
\newcommand{\att}{A_{FB}^{t\bar{t}}}

\documentclass{PoS}

\usepackage{amsmath,bm}
\usepackage{epsfig}

\usepackage{subfigure}

\title{Top quark forward-backward asymmetry at the Tevatron: the electroweak contribution}

\ShortTitle{Top quark $\afb$: the electroweak contribution}

\author{\speaker{Davide Pagani}\\
 Max-Planck-Institut f\"ur Physik
(Werner-Heisenberg-Institut), F\"ohringer Ring 6, 80805 M\"unchen,
Germany\\ 
        E-mail: \email{pagani@mppmu.mpg.de}}        

\abstract{Recent CDF and D\O\  measurements of the top quark forward-backward asymmetry at the Tevatron present deviations between 1$\sigma$ and 2$\sigma$ from QCD predictions and a cut $M_{\ttt} > 450$~GeV on the events increases this deviation to 3.4 $\sigma$ in the CDF measurement. The inclusion of electroweak contributions of $\mathcal{O}(\alpha^2)$ and $\mathcal{O}(\alpha\alphas^{2})$ enlarges the theoretical prediction from QCD by a factor $\sim 1.2$ and diminishes the observed deviations. The calculation method is shortly discussed and numerical results are compared to the experimental data.    }

\FullConference{Proceedings of the Corfu Summer Institute 2011 School and Workshops on Elementary Particle Physics and Gravity\\
		September 4-18, 2011\\
		Corfu, Greece}

\begin{document}
\section{Introduction}
In the experimental analyses at the Tevatron, two different definitions of the forward-backward asymmetry have been used:
\begin{equation}
\label{Afbtt}
  A_{FB}^{t\bar{t}}=
  \frac{\sigma(\Delta y > 0) - \sigma(\Delta y < 0)}{\sigma(\Delta y > 0) + \sigma(\Delta y < 0)}
\end{equation}
and
\begin{equation}
\label{Afblab}
 \alab=\frac{\sigma(y_{t} > 0) - \sigma(y_{t}< 0)}{\sigma(y_{t}> 0) + \sigma(y_{t}< 0)}
\end{equation}
where $\Delta y$ is defined as the difference between the rapidity $y_{t}$ and $y_{\bar{t}}$ of $t$ 
and $\bar{t}$ and the beam axis is oriented in the direction of the proton.
The values obtained by CDF for the inclusive asymmetry ~\cite{Aaltonen:2011kc} are $(A_{FB}^{t\bar{t}}=0.158\pm0.075,~\alab=0.150\pm0.055 )$. $\att$ is compatible with the value obtained by D\O\ $(A_{FB}^{t\bar{t}}=0.196\pm0.065)$ ~\cite{Abazov:2011rq}.\\
All these values are larger than the
Standard Model LO predictions 
 $\att\sim7\%,~\alab\sim5\%$ (see e.g.~\cite{Bernreuther:2010ny}) and imposing a cut $M_{\ttt} > 450$~GeV, the value obtained by CDF $(\att=0.475\pm0.114)$ is at 3.4 $\sigma$ from the prediction at this level of accuracy.    
These results have led to many speculations on the presence of new physics and so a thorough discussion 
of the SM prediction and the corresponding uncertainty is necessary. 
At present, the theoretical accuracy
is limited by the missing calculation of the complete NNLO contribution from QCD to the antisymmetric part of the
$\ttt$ production cross section. 
Besides the strong interaction, also
the electroweak interaction
gives rise to contributions
to the $\ttt$ forward-backward asymmetry.
Although smaller in size, they are not negligible, and 
a careful investigation is an essential ingredient for an improved theoretical prediction.\\
In the following we briefly summarize our calculation and compare numerical results with experimental data. This talk is based essentially on \cite{Hollik:2011ps}.
\section{Outline of the calculation}
\begin{figure}[!h]
\centering
\epsfig{figure=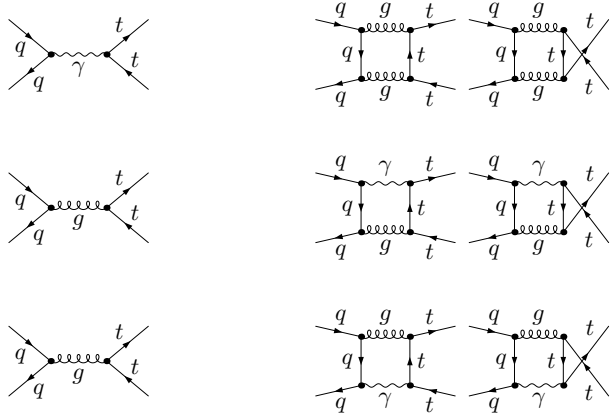,width=8cm}
\caption{Different ways of QED--QCD interference at $\mathcal{O}(\alphas^{2}\alpha)$.}  
\label{bornbox}
\end{figure}

Tree level diagrams of the partonic subprocesses are gluon, photon ad $Z$ s-channel type for $q\bar{q} \rightarrow\ttt$ (Higgs exchange is completely negligible) and s-channel, t-channel and u-channel type for $gg\rightarrow\ttt$.
At leading order the production of $\ttt$ pairs in $p\bar{p}$ collisions originates, via the strong interaction,
from the partonic processes  $q\bar{q}\rightarrow g\rightarrow\ttt$ and $gg\rightarrow\ttt$, which 
yield the $\mathcal{O}(\alphas^{2})$ of the (integrated) cross section, i.e.\ 
the denominator of $\afb$ in \eqref{Afbtt} and \eqref{Afblab}.
Instead the antisymmetric cross section, the numerator of $\afb$, 
starts only at $\mathcal{O}(\alphas^{3})$, so the leading term of the asymmetry involves one loop corrections to $\ttt$ pair production. \\ 
Writing the numerator and the denominator of $\afb$ 
(for either of the definitions  (\ref{Afbtt}) and (\ref{Afblab}))
in powers of $\alphas$  and $\alpha$ we obtain
\begin{eqnarray}\label{powersew}
\nonumber
\afb=\frac{N}{D}&=&\frac{\alpha^{2} \tilde{N}_{0}+\alphas^{3} N_{1}+\alphas^{2}\alpha \tilde{N}_{1}+\alphas^{4} N_{2}+\cdots}{\alpha^{2} \tilde{D}_{0}+\alphas^{2} D_{0}+\alphas^{3} D_{1}+\alphas^{2}\alpha \tilde{D}_{1}+\cdots}=\\
&=&\alphas\frac{N_{1}}{D_{0}}+\alphas^2\frac{(N_{2}-N_{1}D_{1}/D_{0})}{D_{0}}+\alpha\frac{\tilde{N}_{1}}{D_{0}}+\frac{\alpha^{2}}{\alphas^{2}}\frac{\tilde{N}_{0}}{D_{0}}+\cdots
\end{eqnarray}
Only some parts of $N_{2}$ are currently known \cite{Ahrens:2011mw,Ahrens:2011uf} and 
the inclusion of the $N_{1}D_{1}/D_{0}$ term without $N_{2}$ would be 
incomplete, so we have chosen to drop the incomplete $\mathcal{O}(\alphas^{2})$ part, as done in \cite{Kuhn:1998kw}. The inclusion of this term would decrease the asymmetry by about 30\%, which indicates the size of the NLO QCD term that we dropped \footnote{In a very recent paper \cite{Brodsky:2012ik} the expansion of $\afb$ in powers of $\alphas$ has been revisited applying the principle of maximum conformality. In this case $N_{2}$ and also $D_{2}$ seems to be numerically negligible, so the term $N_{1}D_{1}/D_{0}$ can be safely included.}.\\
The remaining terms include $D_{0}$ coming from the leading $\mathcal{O}(\alphas^{2})$ part of the total cross section, $N_{1}$ from the asymmetric part of the NLO QCD corrections to the cross section and $\tilde{N}_0, \tilde{N}_1$ from asymmetric $\mathcal{O}(\alpha^{2}),~\mathcal{O}(\alphas^{2}\alpha)$ parts of the cross section.
In the following we show how these terms arise and how we (re-)evaluated them (for more details see ~\cite{Hollik:2011ps}).\\
The squared terms $|\mathcal{M}_{q\bar{q}\rightarrow g \rightarrow\ttt}|^{2}$ 
and $|\mathcal{M}_{gg\rightarrow \ttt}|^{2}$  yield $D_{0}$ of the LO cross section;
the  $\mathcal{O}(\alpha^{2})$ terms arise from
$|\mathcal{M}_{q\bar{q}\rightarrow\gamma\rightarrow\ttt}+\mathcal{M}_{q\bar{q}\rightarrow  Z \rightarrow\ttt}|^{2}$,
which generate a purely-electroweak antisymmetric differential cross section,
in the parton cms given by
\begin{eqnarray}\label{gamma-z}
\frac{d\sigma_{asym}}{d\cos\theta}=2\pi \alpha^{2}\cos\theta\, \Big(1-\frac{4m_{t}^{2}}{s}\Big)\Big[\kappa\frac{Q_{q}Q_{t}A_{q}A_{t}}{(s-M_{Z}^2)}+2\kappa^{2}A_{q}A_{t}V_{q}V_{t}\frac{s}{(s-M_{Z}^2)^2}\Big], \\
\kappa=\frac{1}{4\sin^{2}(\theta_{W})\cos^{2}(\theta_{W})},\qquad
V_{q}=T^{3}_{q}-2Q_{q}\sin^{2}(\theta_{W}), \qquad A_{q}=T^{3}_{q} , \nonumber
\end{eqnarray}
where $\theta$  is the top-quark scattering angle, $Q_{q}$ and $Q_{t}$ are the charges of the parton $q$ and of the top and $A_{q}$, $A_{t}$ and $V_{q}$, $V_{t}$ are their axial and vectorial couplings to the $Z$ boson. 
In $\afb$ \eqref{powersew} this leads to the term $\tilde{N}_0$. The complementary
symmetric cross section provides the term $\tilde{D}_0$ in the denominator,
which does not contribute in the order under consideration.
Interferences of $q\bar{q}\rightarrow\gamma,Z\rightarrow\ttt$ and
$q\bar{q}\rightarrow g \rightarrow\ttt$ 
are zero  because of the color structure. Basically for $q\bar{q}\rightarrow t\bar{t}$ there are also
 $\mathcal{O}(\alpha)$ $W$-mediated $t$-channel diagrams 
 with $q=d,s,b$, but they are
 strongly suppressed by the CKM matrix or by parton distributions ($q=b$).  \\
 The $\mathcal{O}(\alphas^{3})$ terms that contribute to $N$ arise from four
classes of partonic processes: 
$q\bar{q}\rightarrow\ttt$, $q\bar{q}\rightarrow\ttt g$, 
$qg\rightarrow\ttt q$ and $\bar{q}g\rightarrow\ttt \bar{q}$.
In the first case the origin is the interference of QCD one-loop boxes and
Born amplitudes; the other processes correspond to real-particle emissions.
The box integrals are free of ultraviolet and collinear divergences, but they involve
infrared singularities
which are cancelled after adding the integrated interference 
of initial and final state gluon radiation, the only asymmetric contribution from $q\bar{q}\rightarrow\ttt g$ at $\mathcal{O}(\alphas^{3})$.
$qg\rightarrow\ttt q$ and $\bar{q}g\rightarrow\ttt\bar{q}$ 
yield also contributions to $\afb$, but they are numerically 
not important~\cite{Kuhn:1998kw}.\\
In order to analyze the electroweak $\mathcal{O}(\alphas^{2}\alpha)$ terms, 
it is useful to separate the QED contributions involving photons from the weak
contributions with $Z$ bosons.
In the QED sector we obtain the $\mathcal{O}(\alphas^{2}\alpha)$ contributions
to $N$ from these three classes of partonic processes:
$q\bar{q}\rightarrow\ttt$, $q\bar{q}\rightarrow\ttt g$ and $q\bar{q}\rightarrow\ttt \gamma$.
The first case is the virtual-photon contribution, 
which can be obtained from the QCD analogue, namely the $\mathcal{O}(\alphas^{3})$ 
interference of box and tree-level amplitudes,
by substituting successively each one of the three internal gluons by a photon,
as displayed in Figure~\ref{bornbox}.\\
In a similar way, also the real-radiation processes
$q\bar{q}\rightarrow\ttt g$ and $q\bar{q}\rightarrow\ttt \gamma$ can be
evaluated starting from the result obtained for $q\bar{q}\rightarrow\ttt g$
in the QCD case and  substituting successively each gluon by a photon.\\
The antisymmetric $\mathcal{O}(\alphas^{2}\alpha)$ term from $q\bar{q}\rightarrow\ttt g$ comes from
the interference of $q\bar{q}\rightarrow g \rightarrow \ttt g$
 and $q\bar{q}\rightarrow\gamma\rightarrow\ttt g$, while in the case of
$q\bar{q}\rightarrow\ttt \gamma$ it 
comes from the squared amplitude obtained from $q\bar{q}\rightarrow g \rightarrow \ttt \gamma$ diagrams. 
The essential differences between the calculation of the $\mathcal{O}(\alphas^{3})$
and of QED $\mathcal{O}(\alphas^{2}\alpha)$ terms  
are only the coupling constants and the appearance of the $SU(3)$ generators in the
strong vertices. 
Summing over color in the final state and averaging in the initial state, we find 
that we can relate the QED contribution
of the antisymmetric term $\tilde{N}_1$ in \eqref{powersew} to the
$\mathcal{O}(\alphas^{3})$ QCD term $N_1$ for a given quark species 
$q\bar{q}\rightarrow \ttt(+X)$ in the following way, 
\begin{equation}\label{rqed}
R_{QED}(Q_q) 
             = \frac{\alpha \tilde{N}_{1}^{QED}}{\alpha_s N_{1}}
             = Q_{q}Q_{t}\frac{36}{5}\frac{\alpha}{\alphas} .
\end{equation}
\\Now we consider the weak contribution to $\tilde{N}_1$. 
It can be depicted by the same diagrams as for  
$q\bar{q}\rightarrow\ttt $ and $q\bar{q}\rightarrow\ttt g$ in the QED case, 
but with the photon now substituted by a $Z$ boson, involving massive box
diagrams.
The result cannot be expressed immediately in a simple factorized way. 
We performed the explicit calculation including also the contribution
from real gluon radiation with
numerical integration over the hard gluon part.\\
Also $Z$-boson radiation, $q\bar{q}\rightarrow\ttt Z$, can
contribute at the same order, but it yields 
only a tiny effect of $10^{-5}$ in $\afb$ and thus may be safely neglected.  
The same applies to $u\bar{d}\rightarrow\ttt W^{+}$ as well as to Higgs-boson
radiation.\\
It is important to note that all these partonic subprocesses $p_{1}p_{2}\rightarrow t\bar{t}(+X)$ can be generated with $p_{1}(p_{2})$ coming from the first(second) hadron $h_{1}(h_{2})$ or from $h_{2}(h_{1})$.
Given a kinematic configuration of  $p_{1}p_{2}\rightarrow t\bar{t}(+X)$, if it contributes to $\sigma(Y_{t} > 0)$ in the $h_{1}(h_{2})$ configuration it  contributes with the same partonic weight also to  $\sigma(Y_{t} < 0)$ in the $h_{2}(h_{1})$ configuration. So the total contribution to $\alab$ is non vanishing only if the weights coming from the parton distributions are different, that is if:
\begin{equation}\label{pdfasym}
f_{p_{1},h_{1}}(x_{1})f_{p_{2},h_{2}}(x_{2})\not=f_{p_{1},h_{2}}(x_{1})f_{p_{2},h_{1}}(x_{2})
\end{equation}
where $f_{p_{i},h_{j}}(x_{i})$ is the parton distribution of the parton $p_{i}$ in the hadron $h_{j}$. The same argument applies also to $\att$.\\
At the LHC $h_{1}=h_{2}$ so $\afb$, using definitions \eqref{Afbtt} and \eqref{Afblab}, is equal to zero, at Tevatron $\eqref{pdfasym}$ is not generally true but it can be used to distinguish which subprocesses can give rise to contribution to $\afb$. Only initial states with at least one of the two  $p_{1}$ and $p_{2}$ equal to (anti)quark up or (anti)quark down can produce an asymmetric contribution. This last statement is completely independent on the assumptions made for the partonic calculation, it relies only on the way proton structure is described by partonic distribution functions.
\section{Numerical results}
\begin{figure}[h]
\centering
\subfigure[$\att$\label{thextt}]{
\epsfig{figure=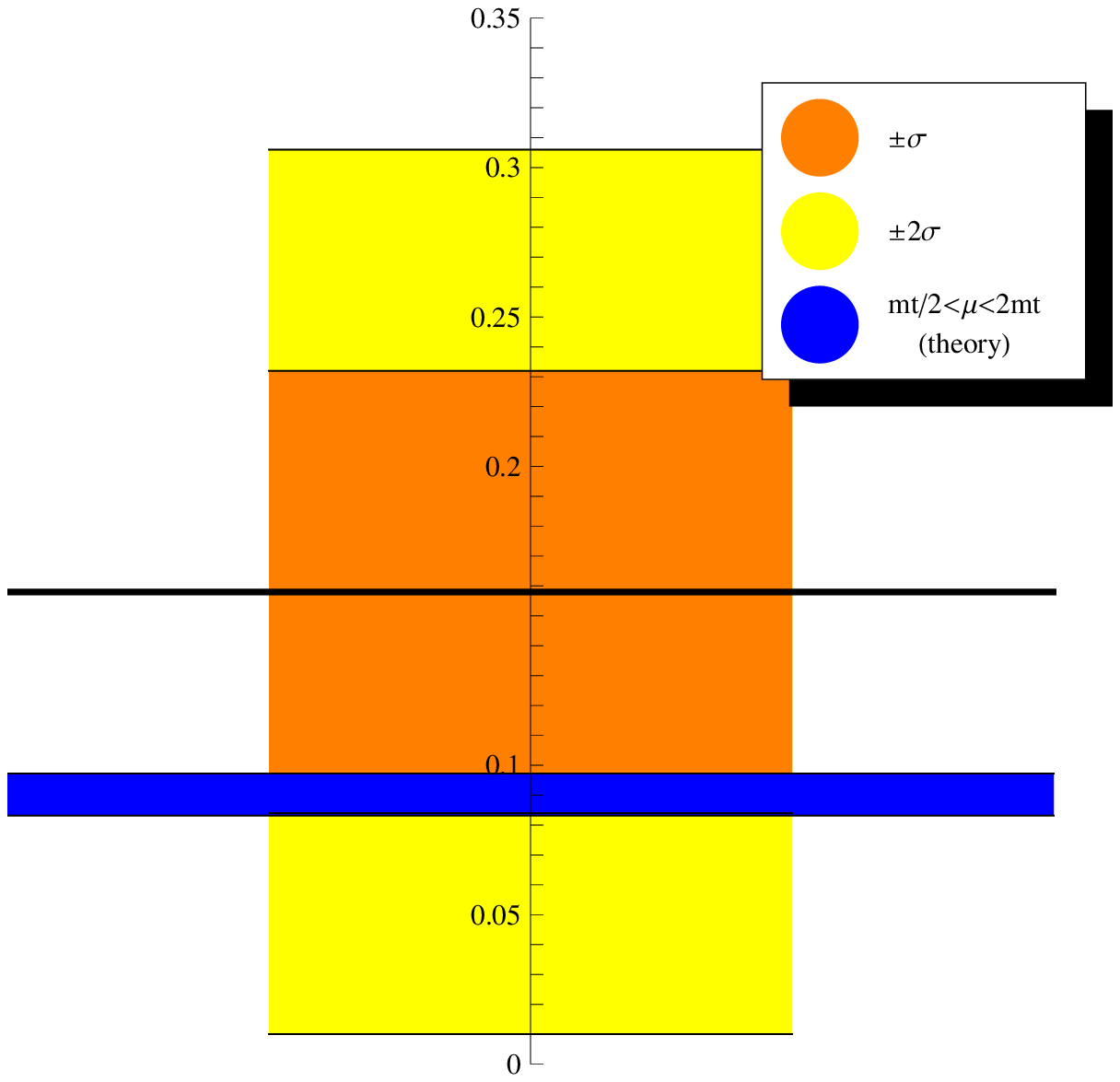,width=7cm}}
\subfigure[$\alab$\label{thexlab}]{
\epsfig{figure=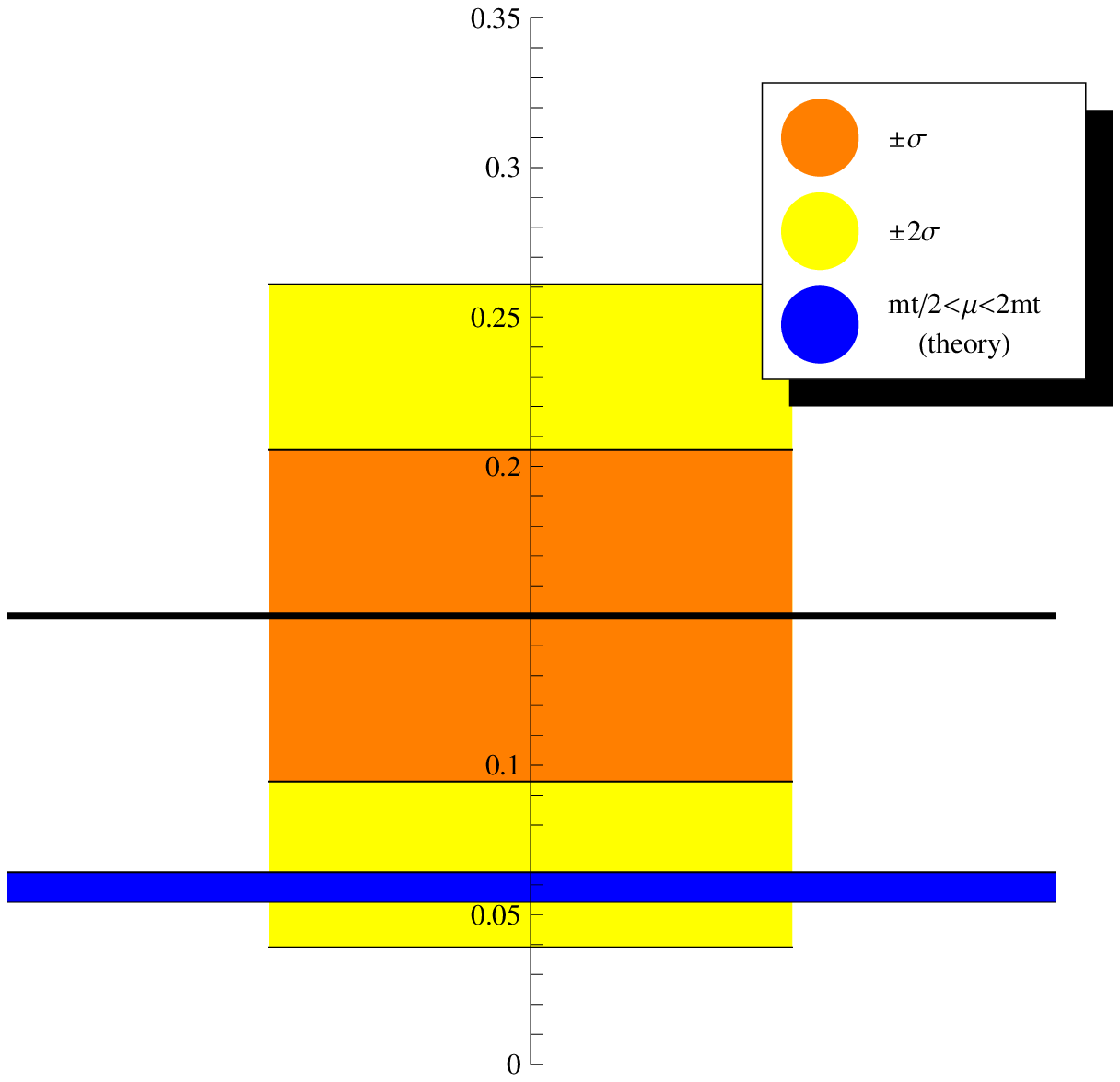,width=7cm}}
\caption{Theoretical prediction (blue) and CDF data (black=central value, orange=$1\sigma$, yellow=$2\sigma$).}
\label{thex}
\end{figure}

According to the argument discussed after \eqref{powersew}, we choose MRST2004QED parton distributions for NLO calculations and MRST2001LO
for LO, 
using thereby $\alphas(\mu)$ of MRST2004QED also for the evaluation of the cross sections at LO
(a similar strategy was employed in~ \cite{Bernreuther:2010ny}).
We used the  same value $\mu$  for the factorization scale and we present
the numerical results with three different choices for the
scale:  $\mu=m_{t}/2,m_{t},2m_{t}$. Other input parameters are taken from \cite{Nakamura:2010zzi}.\\
The various  contributions to the asymmetry of either of the 
two variants $\att$ and $\alab$ are listed 
in  Table~\ref{NNperc}.
\begin{table}[h]
\centering
\subtable[$\att$\label{perctt}]{%
\footnotesize
\begin{tabular}{r|rrr}
\hline
\hline
$\att$		&$\mu=m_{t}/2$	&$\mu=m_{t}$	&$\mu=2m_{t}$\\
\hline
\hline
$\mathcal{O}(\alphas^{3})\quad u\bar{u}$					& 7.01\%		& 6.29\%		& 5.71\%\\
\hline
$\mathcal{O}(\alphas^{3})\quad d\bar{d}$					& 1.16\%		& 1.03\%		&0.92\%\\
\hline
\hline
$\mathcal{O}(\alphas^{2}\alpha)_{QED}\quad u\bar{u}$		& 1.35\%		&1.35\%			&1.35\%	\\
\hline
$\mathcal{O}(\alphas^{2}\alpha)_{QED}\quad d\bar{d}$		& -0.11\%		&-0.11\%		& -0.11\%\\
\hline
\hline
$\mathcal{O}(\alphas^{2}\alpha)_{weak}\quad u\bar{u}$		& 0.16\%		&0.16\%			&  0.16\%	\\
\hline
$\mathcal{O}(\alphas^{2}\alpha)_{weak}\quad d\bar{d}$		& -0.04\%		&-0.04\%		& -0.04\%	\\
\hline
\hline
$\mathcal{O}(\alpha^{2})\quad u\bar{u}$					& 0.18\%		& 0.23\%		& 0.28\%\\
\hline
$\mathcal{O}(\alpha^{2})\quad d\bar{d}$					& 0.02\%		&0.03\%			& 0.03\%\\
\hline
\hline
$\text{tot}\quad p\bar{p}$ 											& 9.72\%		&8.93\%			&8.31\%\\
\hline
\hline
\end{tabular}
}
\subtable[$\alab$\label{perclab}]{%
\footnotesize
\begin{tabular}{r|rrr}
\hline
\hline
$\alab$		&$\mu=m_{t}/2$	&$\mu=m_{t}$	&$\mu=2m_{t}$\\
\hline
\hline
$\mathcal{O}(\alphas^{3})\quad u\bar{u}$					&4.66\%			& 4.19\%		& 3.78\%\\
\hline
$\mathcal{O}(\alphas^{3})\quad d\bar{d}$					& 0.75\%		& 0.66\%		& 0.59\%\\
\hline
\hline
$\mathcal{O}(\alphas^{2}\alpha)_{QED}\quad u\bar{u}$		& 0.90\%		&0.90\%			&0.90\%\\
\hline
$\mathcal{O}(\alphas^{2}\alpha)_{QED}\quad d\bar{d}$		& -0.07\%		&-0.07\%		& -0.07\%\\
\hline
\hline
$\mathcal{O}(\alphas^{2}\alpha)_{weak}\quad u\bar{u}$		& 0.10\%		& 0.10\%		& 0.10\%\\
\hline
$\mathcal{O}(\alphas^{2}\alpha)_{weak}\quad d\bar{d}$		& -0.03\%		&-0.03\%		& -0.03\%\\
\hline
\hline
$\mathcal{O}(\alpha^{2})\quad u\bar{u}$					& 0.11\%		&0.14\%			&  0.17\%\\
\hline
$\mathcal{O}(\alpha^{2})\quad d\bar{d}$					& 0.01\%		&0.02\%			& 0.02\%\\
\hline
\hline
$\text{tot}\quad p\bar{p}$ 											& 6.42\%		&5.92\%			&5.43\%\\
\hline
\hline
\end{tabular}
}
\caption{Different contributions to $\att$ and $\alab$.\label{NNperc}}
\end{table}
The ratio of the total
$\mathcal{O}(\alphas^{2}\alpha)+\mathcal{O}(\alpha^{2})$ and
$\mathcal{O}(\alphas^{3})$ contributions to the numerator $N$
of the asymmetry \eqref{powersew} gives an illustration of the impact of the
electroweak relative to the QCD asymmetry.
The values obtained for $\mu=(m_{t}/2,m_{t},2m_{t})$  
for the two definitions of $\afb$ are
\begin{eqnarray}
\label{rs}
R_{EW}^{\ttt} &=&
\frac{N^{\ttt}_{\mathcal{O}(\alphas^{2}\alpha)+\mathcal{O}(\alpha^{2})}}{N^{\ttt}_{\mathcal{O}(\alphas^{3})}}=(0.190,0.220,0.254), \nonumber\\
R_{EW}^{p\bar{p}} &=&
\frac{N^{p\bar{p}}_{\mathcal{O}(\alphas^{2}\alpha)+\mathcal{O}(\alpha^{2})}}{N^{p\bar{p}}_{\mathcal{O}(\alphas^{3})}}=(0.186,0.218,0.243). 
\end{eqnarray}
\\This shows that the electroweak contribution provides a non-negligible additional part
to the QCD-based antisymmetric cross section with the same overall sign. Thus it
enlarges the Standard Model prediction for the asymmetry
(the electroweak $\mathcal{O}(\alphas^{2}\alpha)$  contribution of
$u\bar{u}\rightarrow\ttt$  to the asymmetry is even bigger than the
$\mathcal{O}(\alphas^{3})$ contribution of $d\bar{d}\rightarrow\ttt$).
\\The recent reevaluation of the mixed EW--QCD contribution 
to $\afb$ in \cite{Kuhn:2011ri} presented values in agreement with our results. 
\\The final result for the two definitions of $\afb$  
can be summarized as follows,
\begin{equation}\label{afbvalues}
\att=(9.7,8.9,8.3)\% , \qquad \alab=(6.4,5.9,5.4)\% . 
\end{equation}
Figure~\ref{thex} displays the theoretical prediction versus the experimental data. 
The SM prediction is almost inside the experimental $1\sigma$ range
for $\att$ and inside the $2\sigma$ range for $\alab$.
It is important to note that the band indicates the scale
variation of the prediction, it does not
account for all the theoretical uncertainties. 
For example, as already noted,
the $\mathcal{O}(\alphas^{4})$ term in $N$ is missing, and
we did not include the $\mathcal{O}(\alphas^{3})$ part in $D$. 
The decrement  by about 30\% obtained by the inclusion of this term can be considered, in a conservative spirit, as an uncertainty from the incomplete NLO calculation for the asymmetry.
\begin{table}[h]
\centering
\subfigure[$\att(M_{\ttt}>450 \text{ GeV})$\label{percminv}]{%
\footnotesize
\begin{tabular}{r|rrr}
\hline
\hline
$\att$		&$\mu=m_{t}/2$	&$\mu=m_{t}$	&$\mu=2m_{t}$\\
\hline
\hline
$\mathcal{O}(\alphas^{3})\quad u\bar{u}$					& 10.13\%		& 9.10\%		& 8.27\%\\
\hline
$\mathcal{O}(\alphas^{3})\quad d\bar{d}$					& 1.44\%		& 1.27\%		&1.14\%\\
\hline
\hline
$\mathcal{O}(\alphas^{2}\alpha)_{QED}\quad u\bar{u}$		& 1.94\%		&1.95\%			&1.96\%	\\
\hline
$\mathcal{O}(\alphas^{2}\alpha)_{QED}\quad d\bar{d}$		& -0.14\%		&-0.14\%		& -0.14\%\\
\hline
\hline
$\mathcal{O}(\alphas^{2}\alpha)_{weak}\quad u\bar{u}$		& 0.28\%		&0.28\%			&  0.28\%	\\
\hline
$\mathcal{O}(\alphas^{2}\alpha)_{weak}\quad d\bar{d}$		& -0.05\%		&-0.05\%		& -0.05\%	\\
\hline
\hline
$\mathcal{O}(\alpha^{2})\quad u\bar{u}$					& 0.26\%		& 0.33\%		& 0.41\%\\
\hline
$\mathcal{O}(\alpha^{2})\quad d\bar{d}$					& 0.03\%		&0.03\%			& 0.04\%\\
\hline
\hline
$\text{tot}\quad p\bar{p}$ 											& 13.90\%		&12.77\%			&11.91\%\\
\hline
\hline
\end{tabular}
}
\caption{Different contributions to $\att(M_{\ttt}>450 \text{ GeV})$. \label{NNpercminvy}}
\end{table}
\begin{figure}[h]
\centering
\subfigure[$\att(M_{\ttt}>450 \text{ GeV})$\label{thextt}]{
\epsfig{figure=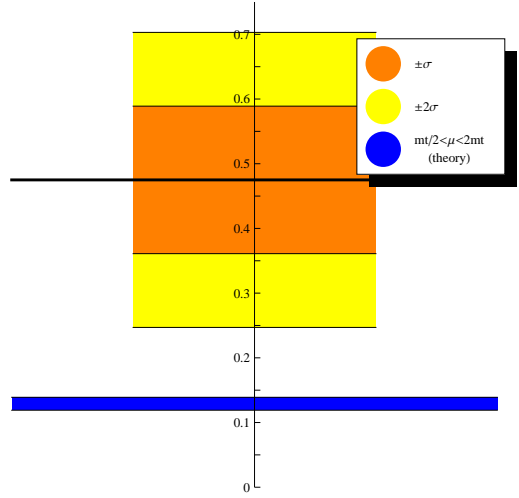,width=7cm}}
\caption{Theoretical prediction (blue) and CDF data (black=central value, orange=$1\sigma$, yellow=$2\sigma$).}
\label{thexminvy}
\end{figure}
\\We have performed our analysis also applying a cut $M_{\ttt}>450 \text{ GeV}$ 
to the $\ttt$ invariant mass. The various contributions to the asymmetry $\att$, as discussed above in the case without cuts, 
are listed
for $M_{\ttt}>450 \text{ GeV}$ in the Table~\ref{NNpercminvy}.
\\The asymmetry with cuts yields
\begin{eqnarray}
\label{afbvaluesminvy}
\att(M_{\ttt}>450 \text{ GeV})=(13.9,12.8,11.9).
\end{eqnarray}
A comparison between Table~\ref{NNpercminvy}(a) and Table~\ref{NNperc}(a) shows 
that the ratio of the QCD contribution to the  $u\bar{u}\rightarrow\ttt$ 
and $d\bar{d}\rightarrow\ttt$ subprocesses is larger with a $M_{\ttt}>450 \text{ GeV}$ cut,
 which leads to a slight increase of $R_{EW}^{\ttt}$:
\begin{equation}\label{rsminvy}
R_{EW}^{\ttt}(M_{\ttt}>450 \text{ GeV})=(0.200,0.232,0.266).
\end{equation}
These values of $R_{EW}^{\ttt}$, however, are not enough to improve the situation, indeed the Standard Model prediction 
is at the $3\sigma$ boundary in case of invariant-mass cut $M_{\ttt}>450 \text{ GeV}$ (see Figure~3 ).
\\In Figure~\ref{D0} the comparison between theoretical prediction and experimental data from D\O\ is shown. The deviation is larger than in the CDF case (Figure~2(a)), but it is important to stress that no statistically significant enhancements have been found by D\O\ for the region according to the cut $M_{\ttt}>450 \text{ GeV}$.
\begin{figure}[!h]
\centering
\subfigure[$\att $\label{thexD0}]{
\epsfig{figure=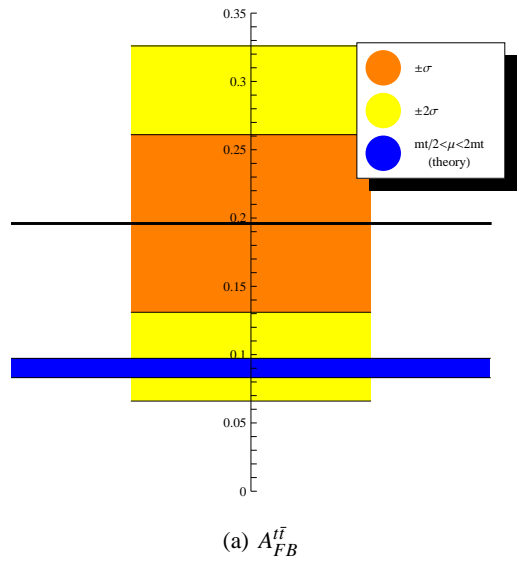,width=7cm}}
\caption{Theoretical prediction (blue) and D\O\ data (black=central value, orange=$1\sigma$, yellow=$2\sigma$).}
\label{D0}
\end{figure}
\section{Conclusions}
The uncertainty of the theoretical prediction for the top quark forward-backward asymmetry at the Tevatron is dominated by the incomplete calculation of the contribution from NNLO QCD corrections to the antisymmetric cross-section. The electroweak contribution is not negligible and increases the LO prediction by a factor $\sim 1.2$, with differences due to the specific definition of the asymmetry and the choice of the renormalization scale.
The main part of these corrections is from QED origin and it can be derived from the LO contribution multiplied by a simple factor depending on the charge of the incoming partons.\\
Electroweak corrections cannot explain the enhancement found by CDF including a cut $M_{\ttt}>450 \text{ GeV}$, but they must not be neglected when the deviation is interpreted as presence of new physics.

\pagebreak

\end{document}